\documentclass[%
 reprint,
unsortedaddress,
nofootinbib,
 amsmath,amssymb,
 aps,
floatfix,
]{revtex4-2}

\usepackage{comment}
\usepackage{graphicx}
\usepackage{dcolumn}
\usepackage{bm}
\usepackage{amsmath}
\usepackage[dvipsnames]{xcolor}

\begin{document}

\preprint{APS/123-QED}

\title{The landscape of instabilities versus shear angle in simulations}

\author{Chloe W. Lindeman}
\email{cwlindeman@jhu.edu}
\author{Sidney R. Nagel}%
\affiliation{%
 Department of Physics and The James Franck and Enrico Fermi Institutes \\ 
 The University  of  Chicago, 
 Chicago,  IL  60637,  USA.\\
\\
}%

\date{\today}%
 
\begin{abstract}
We argue that, in order to appreciate fully how disordered solids fail under sufficiently large strains or thermal noise, a material should be regarded as a network of potentially interacting instabilities occurring along different directions of forcing. We develop a tool for simulations with periodic boundary conditions that allows strains to be applied at a continuously variable angle, $\theta$. This reveals an underlying landscape which consists of lines of instabilities in two dimensions. A modified polar plot provides a visualization of this strain landscape while also allowing a graphic measure of an instability's quadrupolar structure. Many lines persist over a broad angular range of applied strain; some pass through one another; others change continuously with angle or end by smoothly decreasing their magnitudes to zero. Appropriate paths through phase space allow instabilities to be circumvented and hence avoided entirely. We explain this by examining hysterons, \textit{i.e.}, instabilities that undo themselves upon reversing shear direction, and show that as the end of an instability line is approached, the separation between the forward and backward instabilities vanishes.

\end{abstract}

\maketitle

Solids deform when subjected to external stress.  For sufficiently small stress, the deformation is elastic and particles constituting the material return to their initial positions once the perturbation is removed.  Under larger stress, the response can be catastrophic and irreversible. In disordered materials, \textit{e.g.}, jammed packings or glasses, material failure can occur at numerous sites, known as shear-transformation zones, soft spots, or hysterons ~\cite{falk1998dynamics,malandro1999relationships,maloney2004subextensive,maloney2006amorphous, keim2014mechanical, lindeman2024minimal}. In contrast to efforts to predict the locations of individual failure events~\cite{tsamados2009local,manning2011vibrational,cubuk2015identifying,richard2020predicting}, 
we take a broader view of material failure as due to the response of an underlying \textit{network} of instabilities which potentially interact.



There is a large body of simulation work studying material failure under applied stress. Implementations of periodic boundary conditions can allow for varied shear types including  the continuum from triaxial compression to triaxial extension~\cite{lade1975elastoplastic, kraynik1992extensional} (used, for example, in granular rheology and geomechanics~\cite{thornton2010evolution, clemmer2021shear}) and box shape relaxation under applied stress~\cite{parrinello1980crystal, parrinello1981polymorphic, parrinello1982strain}; these are often used to study \textit{global} behavior such as the yield surface or stress behavior under steady state flow. Another class of work examines the local failure events of packings of soft spheres constrained by \textit{fixed} boundaries~\cite{gendelman2015shear,patinet2016connecting,lerbinger2020local, fajardo2026quantifying}. 

Here we combine these approaches by applying shear strain in continuously variable directions, using periodic boundary conditions to minimize  the effects of surfaces, and searching for local rearrangements. We map out lines of instabilities that are encountered as the packing is sheared at different angles and show that going past the first instability reveals a phase-space network of rearrangement events. By examining these networks in detail, we identify features that would be missed using only a fixed strain orientation.

\textit{Transformations for shear at arbitrary $\theta$:} 
Working with periodic boundary conditions requires the  constraint that the box shape must be a parallelogram in $d=2$ or a parallelepiped in $d=3$; the particles that interact with those just inside one surface must be a replica of the properly translated and shifted particles from the opposite boundary. 

\begin{figure}
\includegraphics[width=8.6cm]{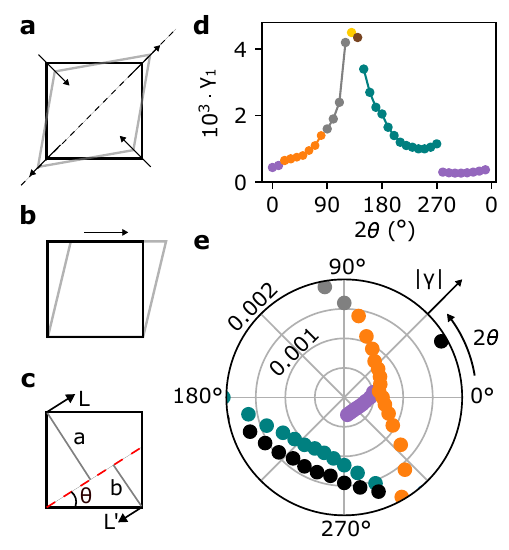}
\caption{
(a) Pure and (b) simple shear along the chosen $\theta=0^{\circ}$ for each case.  (c) Distortion of unit cell to produce simple shear at an angle $\theta$ with respect to the horizontal axis. The dashed line is the line along which there is no displacement. Top left and bottom right vertices are perpendicular distances $b$ and $a$ from that line respectively. Arrows show translation of the vertices parallel to the dashed line, by $L'$ and $L$ in the directions shown. (d) Strain $\gamma_{1} (\theta)$ at which the first rearrangement is encountered when shear is imposed in different directions $2\theta$ ($N=1021$, $d=2$, $\Phi=0.87$). 
(e) The same data as in (d) plotted in polar coordinates and including all rearrangements out to $|\gamma_{\rm{max}}|=0.002$. Plotting all $\gamma_{\alpha} > \gamma_1$ shows the full persistence of the instability lines. Due to the $2\theta$ symmetry of strain, the angular coordinate is $2\theta$. The strain magnitude, $|\gamma|$, is given by the radial distance from the center. Similarly colored points indicate correlated instabilities with $C>0.9$. }
\label{fig:shearder}
\label{fig:rearr}
\end{figure}

In systems in $d$-dimensions with periodic boundaries we define the unit cell in terms of $d$ lattice vectors and require the periodically continued copies to be offset from the original cell by integer multiples of those vectors. Varying the lattice vectors distorts the box and also shifts the neighboring cells so that they tile space smoothly. 
The problem then reduces to defining the appropriate shape of a box that corresponds to pure ($p$) or simple ($s$) shear at an arbitrary angle, $\theta$.

In two dimensions, pure shear corresponds to extension along the positive diagonal, defined as $\theta=0^\circ$, and compression in the orthogonal direction while simple shear corresponds to tilting the cell as in Fig.~\ref{fig:shearder}a,b respectively:
\begin{equation}
\Gamma_{p}(0)=A
\begin{bmatrix}
1 & \frac{\gamma_p}{2} \\
\frac{\gamma_p}{2} & 1 
\end{bmatrix}; ~~~~ \Gamma_{s}(0) = \begin{bmatrix}
1  &  0 \\
\gamma_s & 1
\end{bmatrix}.
\label{eq:zerotheta}
\end{equation} 
In these matrices, each row corresponds to one of the lattice vectors defining the sheared cell.  The prefactor $A=(1 - \frac{\gamma_p^2}{4})^{-1/2}$ for pure shear ensures that cell volume remains constant. For simple shear, $\Gamma_{s}(0)$ corresponds to Lees-Edwards boundary conditions along the horizontal ($\theta=0$) axis~\cite{lees1972computer}. 
The desired transformations can be obtained by conjugating these matrices by a rotation matrix of arbitrary angle, 
\begin{equation}
R_{\theta} = \begin{bmatrix}
\cos{\theta}  &  -\sin{\theta} \\
\sin{\theta} & \cos{\theta}
\end{bmatrix},
\label{eq:rotationmatrix}
\end{equation} 
to find $\Gamma(\theta) = R_{\theta} \Gamma(0) R_{\theta}^{-1}$. This rotated transformation acting on an arbitrary vector $\vec{v}$ is equivalent to rotating $\vec{v}$ by $-\theta$, 
applying the original transformation ($\Gamma(0) R^{-1}_{\theta} \vec{v}$) and then rotating back ($R_{\theta} \Gamma(0) R^{-1}_{\theta} \vec{v}$).
For pure shear, this leads to:

\begin{equation} 
\Gamma_{p}(\theta) =
A
\begin{bmatrix}
1 - \frac{\gamma_{p}}{2} \sin(2\theta) & \frac{\gamma_{p}}{2} \cos(2\theta) \\
\frac{\gamma_{p}}{2} \cos(2\theta) & 1 + \frac{\gamma_{p}}{2} \sin(2\theta)
\end{bmatrix},
\label{eq:2Dthetapure}
\end{equation} 
consistent with~\cite{lindeman2025multi}. For simple shear, we have
\begin{equation} 
\Gamma_{s}(\theta) = 
\begin{bmatrix}
1 - \frac{\gamma_{s}}{2} \sin(2\theta) & -\frac{\gamma_{s}}{2}[1-\cos(2\theta)]\\
\frac{\gamma_{s}}{2} [1+\cos(2\theta)] & 1 + \frac{\gamma_{s}}{2}\sin(2\theta)
\end{bmatrix}.
\label{eq:2Dthetasimple}
\end{equation}  
Since only  $2\theta$ appears, ensuring 
$\Gamma_{p,s}(\theta) = \Gamma_{p,s}(\theta + 180^{\circ})$, we will always label the shear direction in terms of $2\theta$.

Geometrical constructions can give physical intuition for these transformations as shown for the case of simple shear in Fig.~\ref{fig:shearder}c. We require that the dashed line through the origin at angle, $\theta$, remain stationary. We move the vertex $(1, 0)$ by $dL'$ along angle $\theta$ and the vertex $(0,1)$ by $dL$ in the antiparallel direction. A constant shear gradient requires: $dL' = dL (b/a) = dL \tan{\theta}$ where $a$ and $b$ are the two sides of the right triangle as shown.  The shear magnitude is: $\gamma_s = (dL+dL')/(a+b)=dL/\cos{\theta}$. The new vertices are then at: $(1-\gamma_s\sin{\theta}\cos{\theta}, -\gamma_s\sin^2{\theta})$ and ($\gamma_s\cos^2{\theta}, 1+\gamma_s\sin{\theta}\cos{\theta})$ which reduce to 
Eq.~\ref{eq:2Dthetasimple}. A similar construction can be made for pure shear.

These results can be extended to $d=3$, where there are five independent shear deformations. For example, to apply pure shear strain in the $X-Y$ plane:

\begin{equation} 
\Gamma_{p}(\theta) = 
\begin{bmatrix} 
    A(1 - \frac{\gamma_{s}}{2} \sin(2\theta)) 
    & A\frac{\gamma_{p}}{2} \cos(2\theta) & 0 \\
    A\frac{\gamma_{p}}{2} \cos(2\theta) 
    & A(1 + \frac{\gamma_{s}}{2} \sin(2\theta)) & 0
\\   0 &  0 & 1
\end{bmatrix}
\label{eq:3Dthetapure}
\end{equation} 

\textit{Simulations:} We apply pure shear at different angles to packings of frictionless soft discs (in $d=2$) and spheres ($d=3$) using Eqs.~\eqref{eq:2Dthetapure} and~\eqref{eq:3Dthetapure}. As often used in the literature on jamming~\cite{liu2010jamming, van2010jamming}, particles interact via the finite-range, repulsive Hertzian potential:
\begin{equation}
V(r_{ij})=\epsilon(1-r_{ij}/\sigma_{ij})^{5/2}\Theta(\sigma_{ij}-r_{ij}) 
\label{potential}
\end{equation}
where $\epsilon$ sets the energy scale, $ij$ label particles separated by distance $r_{ij}$, $\sigma_{ij}\equiv \sigma_i + \sigma_j$ is the sum of particle radii, and $\Theta(x)$ is the Heaviside function. 
Radii are chosen from a log-normal distribution with polydispersity $P = 0.2$ in $d=2$ and $P = 0.02$ in $d=3$. The particle number, $11\leq N \leq 1021$, and packing fraction, $\Phi$, are as stated.  

Initial configurations are generated with random particle positions and quenched using Fast Inertial Relaxation Engine (FIRE). For each deformation angle, $\theta$ we begin with this same initial configuration and apply athermal quasistatic shear $\Gamma_p(\theta)$ using the pyCudaPacking software package~\cite{morse2014geometric,charbonneau2015jamming}. We choose small increments in $\gamma$, between $10^{-5}$ and $5\times10^{-3}$, and test for rearrangements via reversibility as in~\cite{lindeman2024minimal}: we shear forward one strain step and minimize, then shear back and minimize, comparing the total energy of the packing before and after. This provides an unambiguous metric for determining whether an irreversible rearrangement has occurred. 

\textit{Instability networks: } 
In each direction, we start from the initial configuration and incrementally increase the strain as described above. For each instability, we save the strain at which the $\alpha^{th}$ instability occurs,  $\gamma_{\alpha} (2\theta)$, and the corresponding particle displacements, $\Delta \overrightarrow{r_{\alpha}}(2\theta)$ (an $Nd$-dimensional vector).  Figure~\ref{fig:rearr}d shows, in Cartesian coordinates, an example of $\gamma_{1} (2\theta)$ for a $d=2$ packing of $N=1021$ particles. In some angular regions, \textit{e.g.}, $140^\circ < 2\theta < 270^\circ$, $\gamma_{\alpha}$ varies smoothly with $2\theta$ while elsewhere it abruptly jumps or changes slope.  

\textit{Correlation of instabilities:} In order to compare different rearrangements, the displacements $\Delta \vec{r}$ are first mapped back to the unit square cell by the inverse of the lattice vector matrix: $\Delta \vec{r_0} = \Gamma^{-1}\Delta \vec{r}$.  Following~\cite{patinet2016connecting,lerbinger2020local}, we use these displacements to compare rearrangements $\alpha$ and $\beta$ by calculating the normalized correlation, 
\begin{equation}
C_{\alpha,\beta} = \frac{ \Delta  \overrightarrow{r_{0,\alpha}} \cdot \Delta \overrightarrow{r_{0,\beta}}}{|\Delta \overrightarrow{r_{0,\alpha}}| | \Delta \overrightarrow{r_{0,\beta}}|}.
\label{correlationfunction}
\end{equation} 
Rearrangements with a value of $C_{\alpha,\beta} > 0.9$ are depicted with the same color, confirming that smooth regions in Fig.~\ref{fig:rearr}d correspond to instabilities persisting over a range of $2\theta$; discontinuities reveal a change in the instability. 

\textit{Polar plots:} The same data, $\gamma_{\alpha} (2\theta)$, can also be plotted in polar coordinates as in Fig.~\ref{fig:rearr}e where we have extended the range to include all instabilities out to a maximum strain, $\gamma_{\rm{max}}$. This helps visualize the network of instability lines that determine material failure. The radial distance labels the strain magnitude, $|\gamma|$, while the azimuthal direction labels the extension axis with $2\theta$ plotted so that the periodicity is explicitly manifested.

\textit{Measuring quadrupoles:} Using a $2\theta$ polar plot for pure strain provides a natural tool for testing whether a line of rearrangements has the expected~\cite{eshelby1957determination, chikkadi2011long, maloney2004subextensive,maloney2006amorphous} quadrupolar structure: extensive along one axis, $2\theta_Q$, and compressive in the transverse direction with symmetry of $2\theta$. The triggering strain, $|\gamma(2\theta)|$, will be smallest when the applied strain and inherent deformation directions align. In order to trigger the instability as $2\theta$ varies, the projection onto $2\theta_Q$ should remain constant: $\gamma(2\theta) \cos(2\theta - 2\theta_Q) = \gamma (2\theta_Q)$. For a pure quadrupolar instability, $\gamma(2\theta)$ is thus a straight line on our polar plot. 

Indeed, the curved features seen in Cartesian coordinates (Fig.~\ref{fig:rearr}d), become nearly straight when replotted in polar coordinates (Fig.~\ref{fig:rearr}e). Their straightness measures how close the instabilities are to being ideal quadrupoles in the sense that any deviation from a line indicates an asymmetry in how the instability is triggered. We see indications of this in cases where two rearrangement lines meet, for example in Fig.~\ref{fig:rearr}e near $0^{\circ}$. This suggests that interactions between rearrangements can distort the quadrupolar nature of the instability. 

\textit{Features of instability networks:} Plotting in polar coordinates reveals a number of striking features.  One feature, also found in systems with fixed boundaries~\cite{gendelman2015shear,patinet2016connecting,lerbinger2020local}, is that a single rearrangement can persist over an extended angular region, sometimes greater than $2\theta = 180^{\circ}$.  Such deformations correspond to shear in opposite directions. This large persistence is at odds with a naive picture of shear where one would expect that opposite deformations lead to entirely different responses, as in force chains with an orientation determined by the direction of shear. 

\begin{figure}
\includegraphics[width=8.6cm]{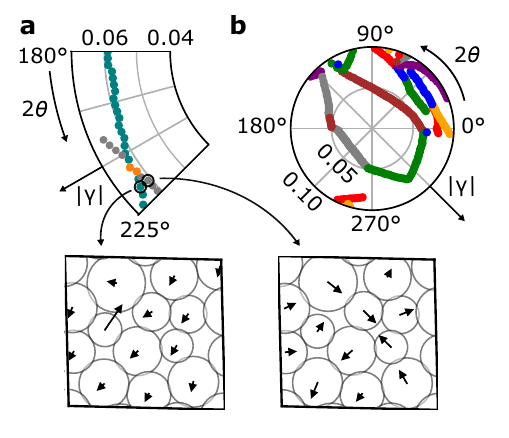}
\caption{Small, $N=13$ systems.  (a) Part of a polar plot showing two rearrangement lines that cross ($d=2, \Phi=0.9$).  The straight instability lines indicate the quadrupolar nature of these instabilities even though that is hard to tell visually from the displacement fields themselves. For two circled rearrangements, spatial structure is shown with arrows indicating the direction and relative magnitudes of displacements (for visualization, packings have not been mapped back to a square box). (b) Polar plot showing the instability network in the X-Y plane of pure shear for a $d=3$ sample. 
} 
\label{fig:N=13}
\end{figure}

Figure~\ref{fig:rearr}e shows that two rearrangement lines can approach one another as $2\theta$ is varied. In a large sample it is intuitive that two different lines can cross one another if they appear in spatially distant regions of the material. In a small sample, however, one might expect that rearrangements must interact strongly. Figure~\ref{fig:N=13}a ($N=13$, $d=2$) shows that this is not necessarily the case: two rearrangements cross largely unimpeded by one another. This is explained by examining their vector displacements, shown below the polar plot --- the two rearrangements are orthogonal.  

\textit{Networks in $d=3$ samples:} Figure~\ref{fig:N=13}b shows a $d=3$ example ($N=13$, $\Phi=0.75$) where we have applied pure shear strain in the $X-Y$ plane. 
Because $d=3$, there are three other shear directions orthogonal to the two shown, yet this slice through phase space shows many of the same features seen in $d=2$: instabilities that persist over large angles and lines of instabilities that collide. 

\begin{figure}
\includegraphics[width=8.6cm]{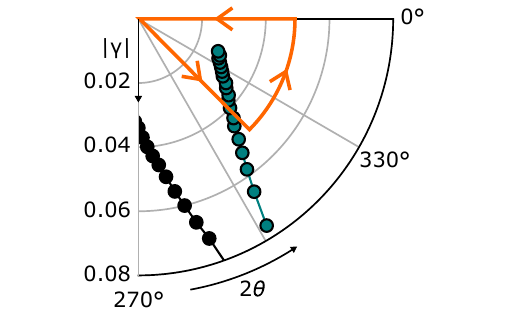}
\caption{An example ($N=11$, $d=2$, $\Phi=0.9$) of a rearrangement that appears to end abruptly (teal rearrangement, end occurs around $2\theta = 340^\circ$). The orange path shows a trajectory that starts at the origin, goes radially outward crossing the teal line of rearrangements, moves azimuthally to $2\theta=360^{\circ}$, and then returns radially inward to the origin. }
 
\label{fig:avoid}
\end{figure}

\textit{Ending of an instability line:}  We illustrate how different rearrangement lines can end using the examples in Fig.~\ref{fig:rearr}e and ~\ref{fig:N=13}. 
In some cases, one rearrangement line runs into another and vanishes as where the purple rearrangement line hits the orange one in Fig.~\ref{fig:rearr}e near $2 \theta = 0^{\circ}$. 
In other cases, a rearrangement line changes abruptly from one rearrangement to another without a large jump in strain as in the transition from orange to grey around $2\theta = 90^{\circ}$ in the same figure. 
Figure~\ref{fig:N=13}(b) shows a case where two rearrangements lines (red and yellow, near $0^{\circ}$) merge into a different rearrangement (blue, starting around $2\theta = 30^{\circ}$). 

\textit{Circumventing an instability:}  A final, surprising case is shown in Fig.~\ref{fig:avoid}: a rearrangement line that ends abruptly without another rearrangement nearby. This suggests that one can entirely avoid an instability by choosing a path that circumvents the corresponding rearrangement line.  This would allow a packing to transform smoothly and reversibly over large regimes of shear. 

To test this, we applied the trajectory shown in  Fig.~\ref{fig:avoid} (solid orange line with arrows) by shearing radially outward such that we cross the teal rearrangement line, then moving azimuthally to $2\theta = 360^\circ$ and finally returning back to the origin. The return path circumvents the rearrangement line. The only rearrangement detected is when moving outward through the teal line, yet the packing returns to its initial configuration upon returning to zero strain. Thus, one may reach identical particle configurations not only via different paths, but via paths both with and without rearrangements.

\begin{figure}
\includegraphics[width=8.9cm]{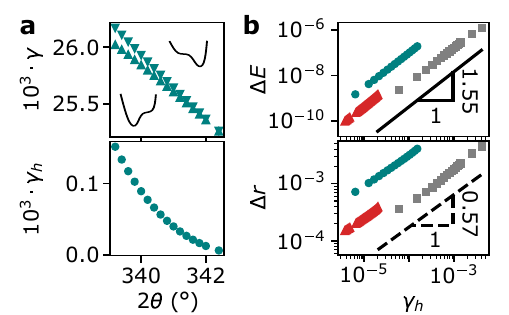}
\caption{ (a) Strain values of rearrangements encountered in both directions $\gamma^+$ and $\gamma^-$ (top) and the difference between them $\gamma_h$ (bottom) versus angle. This data is from the teal-colored instability line shown in Fig.~\ref{fig:avoid}. Cartoons show a schematic of a hysteron as an applied strain switches the system from one potential energy well to another. The switch occurs at different values of strain in the forward and backward directions. (b) Energy drop, $\Delta E$, (top) and particle displacements, $\Delta r$, (bottom) versus hysteresis $\gamma_h$ for each point shown in (b) (teal circles) as well as for two other hysterons measured (grey squares: $N=11$, $\Phi=0.9$; red triangles: $N=17$, $\Phi=0.87$). Average of the three best-fit exponents are shown as guides to the eye ($\Delta E$: $1.55 \pm 0.04$; $\Delta r$: $0.57 \pm 0.05$). 
}
\label{fig:hysteresis}
\end{figure}

\textit{Closing of hysterons:} We can modify the loading protocol to study the behavior of instabilities that belong to the class of rearrangements called \textit{hysterons}: as the strain is increased a rearrangement occurs at a strain $\gamma^+$, but when the system is then sheared radially inward along the same path the rearrangement is undone via a second rearrangement at $\gamma^- < \gamma^+$. This returns the system to its original configuration~\cite{preisach1935magnetische, sethna1993hysteresis, mungan2019networks, keim2020global, terzi2020state, van2021profusion, szulc2022cooperative, shohat2022memory,lindeman2023competition}. Hysterons have played an important role in modeling the memories that can be stored in cyclically sheared solids~\cite{keim2013yielding, regev2013onset, fiocco2013oscillatory, perchikov2014variable, keim2014mechanical, fiocco2014encoding, royer2015precisely, lavrentovich2017period,lindeman2021multiple,lindeman2024minimal}.

We show here that a line of such paired rearrangements can end via vanishing hysteresis: $\gamma_h \equiv \gamma^+ - \gamma^- \rightarrow 0$ as $2\theta$ reaches the end of the rearrangement line.  This is the case for the teal rearrangement line in Fig.~\ref{fig:avoid} that appears to end abruptly. (A second example, for $N=1021$, is shown in Fig.~\ref{fig:rearr}e.) Figure~\ref{fig:hysteresis}a shows both $\gamma^+$ and $\gamma^- $ for the same data. Because $\gamma_h$ quickly becomes small, once rearrangements are found using a fixed step size of $\sim 10^{-5}$, we localize rearrangements to within $10^{-9}$ of the actual strain values $\gamma^+$ and $\gamma^-$ using a bisection algorithm before determining $\gamma_h$,  the energy drop $\Delta E$, and the norm of the particle displacements $\Delta r \equiv \sqrt{\Delta \overrightarrow{r} \cdot \Delta \overrightarrow{r}}$.  

As the hysteron ends, Fig.~\ref{fig:hysteresis}b shows $\Delta E$ and $\Delta r$ both scale with $\gamma_h$; the exponents are consistent with 1.5 for $\Delta E$ and 0.6 for $\Delta r$. 
This is true both for the hysteron shown in Fig.~~\ref{fig:avoid} (teal circles) and for two other examples examined in detail (grey squares: $N=11$, $\Phi=0.9$; red triangles: $N=17$, $\Phi=0.87$). Note that we have focused on the behavior close to the end of each hysteron because the behavior of $\Delta E$ and $\Delta r$ deviate substantially from the smooth behavior far from the endpoints.

A similar, but more noisy, relationship was seen in an ensemble of packings (\textit{i.e.}, different configurations), each measured at a single angle~\cite{lindeman2024minimal}. Varying $\theta$ continuously reveals the scaling is intrinsic to individual rearrangements, with variability between rearrangements in the coefficients to the power law. Moreover, the slow approach of $\gamma_h$ to $0$ as a function of $2\theta$ in Fig.~\ref{fig:hysteresis}a (bottom) explains why there will naturally be an abundance of small hysteresis values as was noted as surprising in~\cite{lindeman2024minimal}.


Hysterons are typically thought of as a double-well energy surface that tilts when a strain is applied as shown schematically in Fig.~\ref{fig:hysteresis}a (top, inset). As indicated in Fig.~\ref{fig:hysteresis}(a), for a range of strains, two stable states exist; the barrier between them produces hysteresis. The continuous vanishing of hysteresis, and disappearance of $\Delta E$ and $\Delta r$ in Fig.~\ref{fig:hysteresis}, suggests that a change in strain direction can cause the two minima to coalesce smoothly. A disappearance of the energy barrier explains how a rearrangement can be circumvented as in Fig.~\ref{fig:avoid} by taking a route where the barrier remains zero.

\textit{Conclusions and future directions:} We have demonstrated how to construct instability networks in simulations with periodic boundary conditions while shear is applied at an arbitrary angle.  This unveils features of material failure that could not be observed using traditional uni-axial pure or simple 
shear: rearrangement lines sometimes cross through one another, other times merge into a single rearrangement, or even disappear as the instability magnitude drops continuously to zero. 

This work opens the door to further studies along several avenues. Though the phase space picture we have developed can also be used to study network solids, suspensions and fluids, we focus here on interesting directions in disordered jammed solids. As shown in Fig.~\ref{fig:N=13}a, it is natural to compare the \textit{shear networks} we have visualized in our modified polar plots with real-space images of particle positions showing the existence of many incipient soft spots~\cite{tsamados2009local,manning2011vibrational,cubuk2015identifying,richard2020predicting}; the spatial proximity of different instabilities can influence how they interact. Moreover, a phase-space map allows one to choose the direction of shear in order to study in detail features like crossing rearrangements or vanishing hysterons. 
Some lines persist over large ranges of $2\theta$ while other do not; an instability with a large persistence angle may be particularly susceptible to thermal excitations since it would be unstable over a large range of fluctuations. If true this would be important in the ongoing hunt for soft spots~\cite{tsamados2009local,manning2011vibrational,cubuk2015identifying,richard2020predicting}.

There is an apparent similarity between the power-law behavior shown in Fig.~\ref{fig:hysteresis}b and features of energy barriers described in previous work~\cite{maloney2006energy}; future work could explore this by examining the relationship between energy barriers, energy drops, and angle of driving for vanishing hysterons. While we have focused on frictionless particles as used in the jamming literature~\cite{liu2010jamming, van2010jamming}, it remains to be seen how frictional interactions alter the phase-space instability networks we have described here.
Finally, it would be interesting to explore correlations between angular persistence and the size of the resulting energy basins to  understand better how materials form memories of the applied deformation.

\section*{Acknowledgments}

We thank Eric Corwin, Spencer Fajardo, Michael Falk, Carl Goodrich, Varda Hagh, Peter Y. Lu, Joshua Mundinger, Muhittin Mungan, Itamar Proccacia, and Damien Vandembroucq for insightful conversations. This work was supported by the US Department of Energy, Office of Science, Basic Energy Sciences, under Grant DE-SC0020972. Computational resources were provided by the NSF MRSEC program NSF-DMR 2011854.

\label{figsi:hysteresis}


\begin{thebibliography}{47}%
\makeatletter
\providecommand \@ifxundefined [1]{%
 \@ifx{#1\undefined}
}%
\providecommand \@ifnum [1]{%
 \ifnum #1\expandafter \@firstoftwo
 \else \expandafter \@secondoftwo
 \fi
}%
\providecommand \@ifx [1]{%
 \ifx #1\expandafter \@firstoftwo
 \else \expandafter \@secondoftwo
 \fi
}%
\providecommand \natexlab [1]{#1}%
\providecommand \enquote  [1]{``#1''}%
\providecommand \bibnamefont  [1]{#1}%
\providecommand \bibfnamefont [1]{#1}%
\providecommand \citenamefont [1]{#1}%
\providecommand \href@noop [0]{\@secondoftwo}%
\providecommand \href [0]{\begingroup \@sanitize@url \@href}%
\providecommand \@href[1]{\@@startlink{#1}\@@href}%
\providecommand \@@href[1]{\endgroup#1\@@endlink}%
\providecommand \@sanitize@url [0]{\catcode `\\12\catcode `\$12\catcode `\&12\catcode `\#12\catcode `\^12\catcode `\_12\catcode `\%12\relax}%
\providecommand \@@startlink[1]{}%
\providecommand \@@endlink[0]{}%
\providecommand \url  [0]{\begingroup\@sanitize@url \@url }%
\providecommand \@url [1]{\endgroup\@href {#1}{\urlprefix }}%
\providecommand \urlprefix  [0]{URL }%
\providecommand \Eprint [0]{\href }%
\providecommand \doibase [0]{https://doi.org/}%
\providecommand \selectlanguage [0]{\@gobble}%
\providecommand \bibinfo  [0]{\@secondoftwo}%
\providecommand \bibfield  [0]{\@secondoftwo}%
\providecommand \translation [1]{[#1]}%
\providecommand \BibitemOpen [0]{}%
\providecommand \bibitemStop [0]{}%
\providecommand \bibitemNoStop [0]{.\EOS\space}%
\providecommand \EOS [0]{\spacefactor3000\relax}%
\providecommand \BibitemShut  [1]{\csname bibitem#1\endcsname}%
\let\auto@bib@innerbib\@empty
\bibitem [{\citenamefont {Falk}\ and\ \citenamefont {Langer}(1998)}]{falk1998dynamics}%
  \BibitemOpen
  \bibfield  {author} {\bibinfo {author} {\bibfnamefont {M.~L.}\ \bibnamefont {Falk}}\ and\ \bibinfo {author} {\bibfnamefont {J.~S.}\ \bibnamefont {Langer}},\ }\bibfield  {title} {\bibinfo {title} {Dynamics of viscoplastic deformation in amorphous solids},\ }\href@noop {} {\bibfield  {journal} {\bibinfo  {journal} {Phys. Rev. E}\ }\textbf {\bibinfo {volume} {57}},\ \bibinfo {pages} {7192} (\bibinfo {year} {1998})}\BibitemShut {NoStop}%
\bibitem [{\citenamefont {Malandro}\ and\ \citenamefont {Lacks}(1999)}]{malandro1999relationships}%
  \BibitemOpen
  \bibfield  {author} {\bibinfo {author} {\bibfnamefont {D.~L.}\ \bibnamefont {Malandro}}\ and\ \bibinfo {author} {\bibfnamefont {D.~J.}\ \bibnamefont {Lacks}},\ }\bibfield  {title} {\bibinfo {title} {Relationships of shear-induced changes in the potential energy landscape to the mechanical properties of ductile glasses},\ }\href@noop {} {\bibfield  {journal} {\bibinfo  {journal} {The Journal of chemical physics}\ }\textbf {\bibinfo {volume} {110}},\ \bibinfo {pages} {4593} (\bibinfo {year} {1999})}\BibitemShut {NoStop}%
\bibitem [{\citenamefont {Maloney}\ and\ \citenamefont {Lemaître}(2004)}]{maloney2004subextensive}%
  \BibitemOpen
  \bibfield  {author} {\bibinfo {author} {\bibfnamefont {C.~E.}\ \bibnamefont {Maloney}}\ and\ \bibinfo {author} {\bibfnamefont {A.}~\bibnamefont {Lemaître}},\ }\bibfield  {title} {\bibinfo {title} {Subextensive scaling in the athermal, quasistatic limit of amorphous matter in plastic shear flow},\ }\href@noop {} {\bibfield  {journal} {\bibinfo  {journal} {Phys. Rev. Lett.}\ }\textbf {\bibinfo {volume} {93}},\ \bibinfo {pages} {016001} (\bibinfo {year} {2004})}\BibitemShut {NoStop}%
\bibitem [{\citenamefont {Maloney}\ and\ \citenamefont {Lemaitre}(2006)}]{maloney2006amorphous}%
  \BibitemOpen
  \bibfield  {author} {\bibinfo {author} {\bibfnamefont {C.~E.}\ \bibnamefont {Maloney}}\ and\ \bibinfo {author} {\bibfnamefont {A.}~\bibnamefont {Lemaitre}},\ }\bibfield  {title} {\bibinfo {title} {Amorphous systems in athermal, quasistatic shear},\ }\href@noop {} {\bibfield  {journal} {\bibinfo  {journal} {Physical Review E—Statistical, Nonlinear, and Soft Matter Physics}\ }\textbf {\bibinfo {volume} {74}},\ \bibinfo {pages} {016118} (\bibinfo {year} {2006})}\BibitemShut {NoStop}%
\bibitem [{\citenamefont {Keim}\ and\ \citenamefont {Arratia}(2014)}]{keim2014mechanical}%
  \BibitemOpen
  \bibfield  {author} {\bibinfo {author} {\bibfnamefont {N.~C.}\ \bibnamefont {Keim}}\ and\ \bibinfo {author} {\bibfnamefont {P.~E.}\ \bibnamefont {Arratia}},\ }\bibfield  {title} {\bibinfo {title} {Mechanical and microscopic properties of the reversible plastic regime in a 2d jammed material},\ }\href@noop {} {\bibfield  {journal} {\bibinfo  {journal} {Phys. Rev. Lett.}\ }\textbf {\bibinfo {volume} {112}},\ \bibinfo {pages} {028302} (\bibinfo {year} {2014})}\BibitemShut {NoStop}%
\bibitem [{\citenamefont {Lindeman}\ and\ \citenamefont {Nagel}(2025)}]{lindeman2024minimal}%
  \BibitemOpen
  \bibfield  {author} {\bibinfo {author} {\bibfnamefont {C.~W.}\ \bibnamefont {Lindeman}}\ and\ \bibinfo {author} {\bibfnamefont {S.~R.}\ \bibnamefont {Nagel}},\ }\bibfield  {title} {\bibinfo {title} {Minimal cyclic behavior in sheared amorphous solids},\ }\href@noop {} {\bibfield  {journal} {\bibinfo  {journal} {New Journal of Physics}\ }\textbf {\bibinfo {volume} {27}},\ \bibinfo {pages} {085001} (\bibinfo {year} {2025})}\BibitemShut {NoStop}%
\bibitem [{\citenamefont {Tsamados}\ \emph {et~al.}(2009)\citenamefont {Tsamados}, \citenamefont {Tanguy}, \citenamefont {Goldenberg},\ and\ \citenamefont {Barrat}}]{tsamados2009local}%
  \BibitemOpen
  \bibfield  {author} {\bibinfo {author} {\bibfnamefont {M.}~\bibnamefont {Tsamados}}, \bibinfo {author} {\bibfnamefont {A.}~\bibnamefont {Tanguy}}, \bibinfo {author} {\bibfnamefont {C.}~\bibnamefont {Goldenberg}},\ and\ \bibinfo {author} {\bibfnamefont {J.-L.}\ \bibnamefont {Barrat}},\ }\bibfield  {title} {\bibinfo {title} {Local elasticity map and plasticity in a model lennard-jones glass},\ }\href@noop {} {\bibfield  {journal} {\bibinfo  {journal} {Phys. Rev. E}\ }\textbf {\bibinfo {volume} {80}},\ \bibinfo {pages} {026112} (\bibinfo {year} {2009})}\BibitemShut {NoStop}%
\bibitem [{\citenamefont {Manning}\ and\ \citenamefont {Liu}(2011)}]{manning2011vibrational}%
  \BibitemOpen
  \bibfield  {author} {\bibinfo {author} {\bibfnamefont {M.~L.}\ \bibnamefont {Manning}}\ and\ \bibinfo {author} {\bibfnamefont {A.~J.}\ \bibnamefont {Liu}},\ }\bibfield  {title} {\bibinfo {title} {Vibrational modes identify soft spots in a sheared disordered packing},\ }\href@noop {} {\bibfield  {journal} {\bibinfo  {journal} {Phys. Rev. Lett.}\ }\textbf {\bibinfo {volume} {107}},\ \bibinfo {pages} {108302} (\bibinfo {year} {2011})}\BibitemShut {NoStop}%
\bibitem [{\citenamefont {Cubuk}\ \emph {et~al.}(2015)\citenamefont {Cubuk}, \citenamefont {Schoenholz}, \citenamefont {Rieser}, \citenamefont {Malone}, \citenamefont {Rottler}, \citenamefont {Durian}, \citenamefont {Kaxiras},\ and\ \citenamefont {Liu}}]{cubuk2015identifying}%
  \BibitemOpen
  \bibfield  {author} {\bibinfo {author} {\bibfnamefont {E.~D.}\ \bibnamefont {Cubuk}}, \bibinfo {author} {\bibfnamefont {S.~S.}\ \bibnamefont {Schoenholz}}, \bibinfo {author} {\bibfnamefont {J.~M.}\ \bibnamefont {Rieser}}, \bibinfo {author} {\bibfnamefont {B.~D.}\ \bibnamefont {Malone}}, \bibinfo {author} {\bibfnamefont {J.}~\bibnamefont {Rottler}}, \bibinfo {author} {\bibfnamefont {D.~J.}\ \bibnamefont {Durian}}, \bibinfo {author} {\bibfnamefont {E.}~\bibnamefont {Kaxiras}},\ and\ \bibinfo {author} {\bibfnamefont {A.~J.}\ \bibnamefont {Liu}},\ }\bibfield  {title} {\bibinfo {title} {Identifying structural flow defects in disordered solids using machine-learning methods},\ }\href@noop {} {\bibfield  {journal} {\bibinfo  {journal} {Phys. Rev. Lett.}\ }\textbf {\bibinfo {volume} {114}},\ \bibinfo {pages} {108001} (\bibinfo {year} {2015})}\BibitemShut {NoStop}%
\bibitem [{\citenamefont {Richard}\ \emph {et~al.}(2020)\citenamefont {Richard}, \citenamefont {Ozawa}, \citenamefont {Patinet}, \citenamefont {Stanifer}, \citenamefont {Shang}, \citenamefont {Ridout}, \citenamefont {Xu}, \citenamefont {Zhang}, \citenamefont {Morse}, \citenamefont {Barrat} \emph {et~al.}}]{richard2020predicting}%
  \BibitemOpen
  \bibfield  {author} {\bibinfo {author} {\bibfnamefont {D.}~\bibnamefont {Richard}}, \bibinfo {author} {\bibfnamefont {M.}~\bibnamefont {Ozawa}}, \bibinfo {author} {\bibfnamefont {S.}~\bibnamefont {Patinet}}, \bibinfo {author} {\bibfnamefont {E.}~\bibnamefont {Stanifer}}, \bibinfo {author} {\bibfnamefont {B.}~\bibnamefont {Shang}}, \bibinfo {author} {\bibfnamefont {S.}~\bibnamefont {Ridout}}, \bibinfo {author} {\bibfnamefont {B.}~\bibnamefont {Xu}}, \bibinfo {author} {\bibfnamefont {G.}~\bibnamefont {Zhang}}, \bibinfo {author} {\bibfnamefont {P.}~\bibnamefont {Morse}}, \bibinfo {author} {\bibfnamefont {J.-L.}\ \bibnamefont {Barrat}}, \emph {et~al.},\ }\bibfield  {title} {\bibinfo {title} {Predicting plasticity in disordered solids from structural indicators},\ }\href@noop {} {\bibfield  {journal} {\bibinfo  {journal} {Phys. Rev. Materials}\ }\textbf {\bibinfo {volume} {4}},\ \bibinfo {pages} {113609} (\bibinfo {year} {2020})}\BibitemShut {NoStop}%
\bibitem [{\citenamefont {Lade}\ and\ \citenamefont {Duncan}(1975)}]{lade1975elastoplastic}%
  \BibitemOpen
  \bibfield  {author} {\bibinfo {author} {\bibfnamefont {P.~V.}\ \bibnamefont {Lade}}\ and\ \bibinfo {author} {\bibfnamefont {J.~M.}\ \bibnamefont {Duncan}},\ }\bibfield  {title} {\bibinfo {title} {Elastoplastic stress-strain theory for cohesionless soil},\ }\href@noop {} {\bibfield  {journal} {\bibinfo  {journal} {Journal of the Geotechnical Engineering Division}\ }\textbf {\bibinfo {volume} {101}},\ \bibinfo {pages} {1037} (\bibinfo {year} {1975})}\BibitemShut {NoStop}%
\bibitem [{\citenamefont {Kraynik}\ and\ \citenamefont {Reinelt}(1992)}]{kraynik1992extensional}%
  \BibitemOpen
  \bibfield  {author} {\bibinfo {author} {\bibfnamefont {A.}~\bibnamefont {Kraynik}}\ and\ \bibinfo {author} {\bibfnamefont {D.}~\bibnamefont {Reinelt}},\ }\bibfield  {title} {\bibinfo {title} {Extensional motions of spatially periodic lattices},\ }\href@noop {} {\bibfield  {journal} {\bibinfo  {journal} {International journal of multiphase flow}\ }\textbf {\bibinfo {volume} {18}},\ \bibinfo {pages} {1045} (\bibinfo {year} {1992})}\BibitemShut {NoStop}%
\bibitem [{\citenamefont {Thornton}\ and\ \citenamefont {Zhang}(2010)}]{thornton2010evolution}%
  \BibitemOpen
  \bibfield  {author} {\bibinfo {author} {\bibfnamefont {C.}~\bibnamefont {Thornton}}\ and\ \bibinfo {author} {\bibfnamefont {L.}~\bibnamefont {Zhang}},\ }\bibfield  {title} {\bibinfo {title} {On the evolution of stress and microstructure during general 3d deviatoric straining of granular media},\ }\href@noop {} {\bibfield  {journal} {\bibinfo  {journal} {G{\'e}otechnique}\ }\textbf {\bibinfo {volume} {60}},\ \bibinfo {pages} {333} (\bibinfo {year} {2010})}\BibitemShut {NoStop}%
\bibitem [{\citenamefont {Clemmer}\ \emph {et~al.}(2021)\citenamefont {Clemmer}, \citenamefont {Srivastava}, \citenamefont {Grest},\ and\ \citenamefont {Lechman}}]{clemmer2021shear}%
  \BibitemOpen
  \bibfield  {author} {\bibinfo {author} {\bibfnamefont {J.~T.}\ \bibnamefont {Clemmer}}, \bibinfo {author} {\bibfnamefont {I.}~\bibnamefont {Srivastava}}, \bibinfo {author} {\bibfnamefont {G.~S.}\ \bibnamefont {Grest}},\ and\ \bibinfo {author} {\bibfnamefont {J.~B.}\ \bibnamefont {Lechman}},\ }\bibfield  {title} {\bibinfo {title} {Shear is not always simple: Rate-dependent effects of flow type on granular rheology},\ }\href@noop {} {\bibfield  {journal} {\bibinfo  {journal} {Physical review letters}\ }\textbf {\bibinfo {volume} {127}},\ \bibinfo {pages} {268003} (\bibinfo {year} {2021})}\BibitemShut {NoStop}%
\bibitem [{\citenamefont {Parrinello}\ and\ \citenamefont {Rahman}(1980)}]{parrinello1980crystal}%
  \BibitemOpen
  \bibfield  {author} {\bibinfo {author} {\bibfnamefont {M.}~\bibnamefont {Parrinello}}\ and\ \bibinfo {author} {\bibfnamefont {A.}~\bibnamefont {Rahman}},\ }\bibfield  {title} {\bibinfo {title} {Crystal structure and pair potentials: A molecular-dynamics study},\ }\href@noop {} {\bibfield  {journal} {\bibinfo  {journal} {Phys. Rev. Lett.}\ }\textbf {\bibinfo {volume} {45}},\ \bibinfo {pages} {1196} (\bibinfo {year} {1980})}\BibitemShut {NoStop}%
\bibitem [{\citenamefont {Parrinello}\ and\ \citenamefont {Rahman}(1981)}]{parrinello1981polymorphic}%
  \BibitemOpen
  \bibfield  {author} {\bibinfo {author} {\bibfnamefont {M.}~\bibnamefont {Parrinello}}\ and\ \bibinfo {author} {\bibfnamefont {A.}~\bibnamefont {Rahman}},\ }\bibfield  {title} {\bibinfo {title} {Polymorphic transitions in single crystals: A new molecular dynamics method},\ }\href@noop {} {\bibfield  {journal} {\bibinfo  {journal} {J. Appl. Phys.}\ }\textbf {\bibinfo {volume} {52}},\ \bibinfo {pages} {7182} (\bibinfo {year} {1981})}\BibitemShut {NoStop}%
\bibitem [{\citenamefont {Parrinello}\ and\ \citenamefont {Rahman}(1982)}]{parrinello1982strain}%
  \BibitemOpen
  \bibfield  {author} {\bibinfo {author} {\bibfnamefont {M.}~\bibnamefont {Parrinello}}\ and\ \bibinfo {author} {\bibfnamefont {A.}~\bibnamefont {Rahman}},\ }\bibfield  {title} {\bibinfo {title} {Strain fluctuations and elastic constants},\ }\href@noop {} {\bibfield  {journal} {\bibinfo  {journal} {J. Chem. Phys.}\ }\textbf {\bibinfo {volume} {76}},\ \bibinfo {pages} {2662} (\bibinfo {year} {1982})}\BibitemShut {NoStop}%
\bibitem [{\citenamefont {Gendelman}\ \emph {et~al.}(2015)\citenamefont {Gendelman}, \citenamefont {Jaiswal}, \citenamefont {Procaccia}, \citenamefont {Gupta},\ and\ \citenamefont {Zylberg}}]{gendelman2015shear}%
  \BibitemOpen
  \bibfield  {author} {\bibinfo {author} {\bibfnamefont {O.}~\bibnamefont {Gendelman}}, \bibinfo {author} {\bibfnamefont {P.~K.}\ \bibnamefont {Jaiswal}}, \bibinfo {author} {\bibfnamefont {I.}~\bibnamefont {Procaccia}}, \bibinfo {author} {\bibfnamefont {B.~S.}\ \bibnamefont {Gupta}},\ and\ \bibinfo {author} {\bibfnamefont {J.}~\bibnamefont {Zylberg}},\ }\bibfield  {title} {\bibinfo {title} {Shear transformation zones: State determined or protocol dependent?},\ }\href@noop {} {\bibfield  {journal} {\bibinfo  {journal} {Europhysics Letters}\ }\textbf {\bibinfo {volume} {109}},\ \bibinfo {pages} {16002} (\bibinfo {year} {2015})}\BibitemShut {NoStop}%
\bibitem [{\citenamefont {Patinet}\ \emph {et~al.}(2016)\citenamefont {Patinet}, \citenamefont {Vandembroucq},\ and\ \citenamefont {Falk}}]{patinet2016connecting}%
  \BibitemOpen
  \bibfield  {author} {\bibinfo {author} {\bibfnamefont {S.}~\bibnamefont {Patinet}}, \bibinfo {author} {\bibfnamefont {D.}~\bibnamefont {Vandembroucq}},\ and\ \bibinfo {author} {\bibfnamefont {M.~L.}\ \bibnamefont {Falk}},\ }\bibfield  {title} {\bibinfo {title} {Connecting local yield stresses with plastic activity in amorphous solids},\ }\href@noop {} {\bibfield  {journal} {\bibinfo  {journal} {Phys. Rev. Lett.}\ }\textbf {\bibinfo {volume} {117}},\ \bibinfo {pages} {045501} (\bibinfo {year} {2016})}\BibitemShut {NoStop}%
\bibitem [{\citenamefont {Lerbinger}(2020)}]{lerbinger2020local}%
  \BibitemOpen
  \bibfield  {author} {\bibinfo {author} {\bibfnamefont {M.}~\bibnamefont {Lerbinger}},\ }\emph {\bibinfo {title} {Local shear rearrangements in glassy systems: from micromechanics to structural relaxation in supercooled liquids}},\ \href@noop {} {Ph.D. thesis},\ \bibinfo  {school} {Universit{\'e} Paris sciences et lettres} (\bibinfo {year} {2020})\BibitemShut {NoStop}%
\bibitem [{\citenamefont {Fajardo}\ \emph {et~al.}(2026)\citenamefont {Fajardo}, \citenamefont {Desmarchelier}, \citenamefont {Patinet},\ and\ \citenamefont {Falk}}]{fajardo2026quantifying}%
  \BibitemOpen
  \bibfield  {author} {\bibinfo {author} {\bibfnamefont {S.}~\bibnamefont {Fajardo}}, \bibinfo {author} {\bibfnamefont {P.}~\bibnamefont {Desmarchelier}}, \bibinfo {author} {\bibfnamefont {S.}~\bibnamefont {Patinet}},\ and\ \bibinfo {author} {\bibfnamefont {M.~L.}\ \bibnamefont {Falk}},\ }\bibfield  {title} {\bibinfo {title} {Quantifying the features of an amorphous solid's local yield surface},\ }\href@noop {} {\bibfield  {journal} {\bibinfo  {journal} {arXiv preprint arXiv:2603.16905}\ } (\bibinfo {year} {2026})}\BibitemShut {NoStop}%
\bibitem [{\citenamefont {Lees}\ and\ \citenamefont {Edwards}(1972)}]{lees1972computer}%
  \BibitemOpen
  \bibfield  {author} {\bibinfo {author} {\bibfnamefont {A.~W.}\ \bibnamefont {Lees}}\ and\ \bibinfo {author} {\bibfnamefont {S.~F.}\ \bibnamefont {Edwards}},\ }\bibfield  {title} {\bibinfo {title} {The computer study of transport processes under extreme conditions},\ }\href@noop {} {\bibfield  {journal} {\bibinfo  {journal} {Journal of Physics C: Solid State Physics}\ }\textbf {\bibinfo {volume} {5}},\ \bibinfo {pages} {1921} (\bibinfo {year} {1972})}\BibitemShut {NoStop}%
\bibitem [{\citenamefont {Lindeman}(2025)}]{lindeman2025multi}%
  \BibitemOpen
  \bibfield  {author} {\bibinfo {author} {\bibfnamefont {C.~W.}\ \bibnamefont {Lindeman}},\ }\bibfield  {title} {\bibinfo {title} {Multi-dimensional memory in low-friction granular materials},\ }\href@noop {} {\bibfield  {journal} {\bibinfo  {journal} {Soft Matter}\ }\textbf {\bibinfo {volume} {21}},\ \bibinfo {pages} {4890} (\bibinfo {year} {2025})}\BibitemShut {NoStop}%
\bibitem [{\citenamefont {Liu}\ and\ \citenamefont {Nagel}(2010)}]{liu2010jamming}%
  \BibitemOpen
  \bibfield  {author} {\bibinfo {author} {\bibfnamefont {A.~J.}\ \bibnamefont {Liu}}\ and\ \bibinfo {author} {\bibfnamefont {S.~R.}\ \bibnamefont {Nagel}},\ }\bibfield  {title} {\bibinfo {title} {The jamming transition and the marginally jammed solid},\ }\href@noop {} {\bibfield  {journal} {\bibinfo  {journal} {Annu. Rev. Condens. Matter Phys.}\ }\textbf {\bibinfo {volume} {1}},\ \bibinfo {pages} {347} (\bibinfo {year} {2010})}\BibitemShut {NoStop}%
\bibitem [{\citenamefont {van Hecke}(2010)}]{van2010jamming}%
  \BibitemOpen
  \bibfield  {author} {\bibinfo {author} {\bibfnamefont {M.}~\bibnamefont {van Hecke}},\ }\bibfield  {title} {\bibinfo {title} {Jamming of soft particles: geometry, mechanics, scaling and isostaticity},\ }\href@noop {} {\bibfield  {journal} {\bibinfo  {journal} {Journal of Physics: Condensed Matter}\ }\textbf {\bibinfo {volume} {22}},\ \bibinfo {pages} {033101} (\bibinfo {year} {2010})}\BibitemShut {NoStop}%
\bibitem [{\citenamefont {Morse}\ and\ \citenamefont {Corwin}(2014)}]{morse2014geometric}%
  \BibitemOpen
  \bibfield  {author} {\bibinfo {author} {\bibfnamefont {P.~K.}\ \bibnamefont {Morse}}\ and\ \bibinfo {author} {\bibfnamefont {E.~I.}\ \bibnamefont {Corwin}},\ }\bibfield  {title} {\bibinfo {title} {Geometric signatures of jamming in the mechanical vacuum},\ }\href@noop {} {\bibfield  {journal} {\bibinfo  {journal} {Phys. Rev. Lett.}\ }\textbf {\bibinfo {volume} {112}},\ \bibinfo {pages} {115701} (\bibinfo {year} {2014})}\BibitemShut {NoStop}%
\bibitem [{\citenamefont {Charbonneau}\ \emph {et~al.}(2015)\citenamefont {Charbonneau}, \citenamefont {Corwin}, \citenamefont {Parisi},\ and\ \citenamefont {Zamponi}}]{charbonneau2015jamming}%
  \BibitemOpen
  \bibfield  {author} {\bibinfo {author} {\bibfnamefont {P.}~\bibnamefont {Charbonneau}}, \bibinfo {author} {\bibfnamefont {E.~I.}\ \bibnamefont {Corwin}}, \bibinfo {author} {\bibfnamefont {G.}~\bibnamefont {Parisi}},\ and\ \bibinfo {author} {\bibfnamefont {F.}~\bibnamefont {Zamponi}},\ }\bibfield  {title} {\bibinfo {title} {Jamming criticality revealed by removing localized buckling excitations},\ }\href@noop {} {\bibfield  {journal} {\bibinfo  {journal} {Phys. Rev. Lett.}\ }\textbf {\bibinfo {volume} {114}},\ \bibinfo {pages} {125504} (\bibinfo {year} {2015})}\BibitemShut {NoStop}%
\bibitem [{\citenamefont {Eshelby}(1957)}]{eshelby1957determination}%
  \BibitemOpen
  \bibfield  {author} {\bibinfo {author} {\bibfnamefont {J.~D.}\ \bibnamefont {Eshelby}},\ }\bibfield  {title} {\bibinfo {title} {The determination of the elastic field of an ellipsoidal inclusion, and related problems},\ }\href@noop {} {\bibfield  {journal} {\bibinfo  {journal} {Proceedings of the Royal Society of London. Series A. Mathematical and physical sciences}\ }\textbf {\bibinfo {volume} {241}},\ \bibinfo {pages} {376} (\bibinfo {year} {1957})}\BibitemShut {NoStop}%
\bibitem [{\citenamefont {Chikkadi}\ \emph {et~al.}(2011)\citenamefont {Chikkadi}, \citenamefont {Wegdam}, \citenamefont {Bonn}, \citenamefont {Nienhuis},\ and\ \citenamefont {Schall}}]{chikkadi2011long}%
  \BibitemOpen
  \bibfield  {author} {\bibinfo {author} {\bibfnamefont {V.}~\bibnamefont {Chikkadi}}, \bibinfo {author} {\bibfnamefont {G.}~\bibnamefont {Wegdam}}, \bibinfo {author} {\bibfnamefont {D.}~\bibnamefont {Bonn}}, \bibinfo {author} {\bibfnamefont {B.}~\bibnamefont {Nienhuis}},\ and\ \bibinfo {author} {\bibfnamefont {P.}~\bibnamefont {Schall}},\ }\bibfield  {title} {\bibinfo {title} {Long-range strain correlations in sheared colloidal glasses},\ }\href@noop {} {\bibfield  {journal} {\bibinfo  {journal} {Phys. Rev. Lett.}\ }\textbf {\bibinfo {volume} {107}},\ \bibinfo {pages} {198303} (\bibinfo {year} {2011})}\BibitemShut {NoStop}%
\bibitem [{\citenamefont {Preisach}(1935)}]{preisach1935magnetische}%
  \BibitemOpen
  \bibfield  {author} {\bibinfo {author} {\bibfnamefont {F.}~\bibnamefont {Preisach}},\ }\bibfield  {title} {\bibinfo {title} {{\"U}ber die magnetische nachwirkung},\ }\href@noop {} {\bibfield  {journal} {\bibinfo  {journal} {Zeitschrift f{\"u}r physik}\ }\textbf {\bibinfo {volume} {94}},\ \bibinfo {pages} {277} (\bibinfo {year} {1935})}\BibitemShut {NoStop}%
\bibitem [{\citenamefont {Sethna}\ \emph {et~al.}(1993)\citenamefont {Sethna}, \citenamefont {Dahmen}, \citenamefont {Kartha}, \citenamefont {Krumhansl}, \citenamefont {Roberts},\ and\ \citenamefont {Shore}}]{sethna1993hysteresis}%
  \BibitemOpen
  \bibfield  {author} {\bibinfo {author} {\bibfnamefont {J.~P.}\ \bibnamefont {Sethna}}, \bibinfo {author} {\bibfnamefont {K.}~\bibnamefont {Dahmen}}, \bibinfo {author} {\bibfnamefont {S.}~\bibnamefont {Kartha}}, \bibinfo {author} {\bibfnamefont {J.~A.}\ \bibnamefont {Krumhansl}}, \bibinfo {author} {\bibfnamefont {B.~W.}\ \bibnamefont {Roberts}},\ and\ \bibinfo {author} {\bibfnamefont {J.~D.}\ \bibnamefont {Shore}},\ }\bibfield  {title} {\bibinfo {title} {Hysteresis and hierarchies: Dynamics of disorder-driven first-order phase transformations},\ }\href@noop {} {\bibfield  {journal} {\bibinfo  {journal} {Phys. Rev. Lett.}\ }\textbf {\bibinfo {volume} {70}},\ \bibinfo {pages} {3347} (\bibinfo {year} {1993})}\BibitemShut {NoStop}%
\bibitem [{\citenamefont {Mungan}\ \emph {et~al.}(2019)\citenamefont {Mungan}, \citenamefont {Sastry}, \citenamefont {Dahmen},\ and\ \citenamefont {Regev}}]{mungan2019networks}%
  \BibitemOpen
  \bibfield  {author} {\bibinfo {author} {\bibfnamefont {M.}~\bibnamefont {Mungan}}, \bibinfo {author} {\bibfnamefont {S.}~\bibnamefont {Sastry}}, \bibinfo {author} {\bibfnamefont {K.}~\bibnamefont {Dahmen}},\ and\ \bibinfo {author} {\bibfnamefont {I.}~\bibnamefont {Regev}},\ }\bibfield  {title} {\bibinfo {title} {Networks and hierarchies: How amorphous materials learn to remember},\ }\href@noop {} {\bibfield  {journal} {\bibinfo  {journal} {Phys. Rev. Lett.}\ }\textbf {\bibinfo {volume} {123}},\ \bibinfo {pages} {178002} (\bibinfo {year} {2019})}\BibitemShut {NoStop}%
\bibitem [{\citenamefont {Keim}\ \emph {et~al.}(2020)\citenamefont {Keim}, \citenamefont {Hass}, \citenamefont {Kroger},\ and\ \citenamefont {Wieker}}]{keim2020global}%
  \BibitemOpen
  \bibfield  {author} {\bibinfo {author} {\bibfnamefont {N.~C.}\ \bibnamefont {Keim}}, \bibinfo {author} {\bibfnamefont {J.}~\bibnamefont {Hass}}, \bibinfo {author} {\bibfnamefont {B.}~\bibnamefont {Kroger}},\ and\ \bibinfo {author} {\bibfnamefont {D.}~\bibnamefont {Wieker}},\ }\bibfield  {title} {\bibinfo {title} {Global memory from local hysteresis in an amorphous solid},\ }\href@noop {} {\bibfield  {journal} {\bibinfo  {journal} {Phys. Rev. Research}\ }\textbf {\bibinfo {volume} {2}},\ \bibinfo {pages} {012004} (\bibinfo {year} {2020})}\BibitemShut {NoStop}%
\bibitem [{\citenamefont {Terzi}\ and\ \citenamefont {Mungan}(2020)}]{terzi2020state}%
  \BibitemOpen
  \bibfield  {author} {\bibinfo {author} {\bibfnamefont {M.~M.}\ \bibnamefont {Terzi}}\ and\ \bibinfo {author} {\bibfnamefont {M.}~\bibnamefont {Mungan}},\ }\bibfield  {title} {\bibinfo {title} {State transition graph of the preisach model and the role of return-point memory},\ }\href@noop {} {\bibfield  {journal} {\bibinfo  {journal} {Phys. Rev. E}\ }\textbf {\bibinfo {volume} {102}},\ \bibinfo {pages} {012122} (\bibinfo {year} {2020})}\BibitemShut {NoStop}%
\bibitem [{\citenamefont {van Hecke}(2021)}]{van2021profusion}%
  \BibitemOpen
  \bibfield  {author} {\bibinfo {author} {\bibfnamefont {M.}~\bibnamefont {van Hecke}},\ }\bibfield  {title} {\bibinfo {title} {Profusion of transition pathways for interacting hysterons},\ }\href@noop {} {\bibfield  {journal} {\bibinfo  {journal} {Phys. Rev. E}\ }\textbf {\bibinfo {volume} {104}},\ \bibinfo {pages} {054608} (\bibinfo {year} {2021})}\BibitemShut {NoStop}%
\bibitem [{\citenamefont {Szulc}\ \emph {et~al.}(2022)\citenamefont {Szulc}, \citenamefont {Mungan},\ and\ \citenamefont {Regev}}]{szulc2022cooperative}%
  \BibitemOpen
  \bibfield  {author} {\bibinfo {author} {\bibfnamefont {A.}~\bibnamefont {Szulc}}, \bibinfo {author} {\bibfnamefont {M.}~\bibnamefont {Mungan}},\ and\ \bibinfo {author} {\bibfnamefont {I.}~\bibnamefont {Regev}},\ }\bibfield  {title} {\bibinfo {title} {Cooperative effects driving the multi-periodic dynamics of cyclically sheared amorphous solids},\ }\href@noop {} {\bibfield  {journal} {\bibinfo  {journal} {J. Chem. Phys.}\ }\textbf {\bibinfo {volume} {156}} (\bibinfo {year} {2022})}\BibitemShut {NoStop}%
\bibitem [{\citenamefont {Shohat}\ \emph {et~al.}(2022)\citenamefont {Shohat}, \citenamefont {Hexner},\ and\ \citenamefont {Lahini}}]{shohat2022memory}%
  \BibitemOpen
  \bibfield  {author} {\bibinfo {author} {\bibfnamefont {D.}~\bibnamefont {Shohat}}, \bibinfo {author} {\bibfnamefont {D.}~\bibnamefont {Hexner}},\ and\ \bibinfo {author} {\bibfnamefont {Y.}~\bibnamefont {Lahini}},\ }\bibfield  {title} {\bibinfo {title} {Memory from coupled instabilities in unfolded crumpled sheets},\ }\href@noop {} {\bibfield  {journal} {\bibinfo  {journal} {Proc. Nat. Acad. Sci.}\ }\textbf {\bibinfo {volume} {119}},\ \bibinfo {pages} {e2200028119} (\bibinfo {year} {2022})}\BibitemShut {NoStop}%
\bibitem [{\citenamefont {Lindeman}\ \emph {et~al.}(2023)\citenamefont {Lindeman}, \citenamefont {Hagh}, \citenamefont {Ip},\ and\ \citenamefont {Nagel}}]{lindeman2023competition}%
  \BibitemOpen
  \bibfield  {author} {\bibinfo {author} {\bibfnamefont {C.~W.}\ \bibnamefont {Lindeman}}, \bibinfo {author} {\bibfnamefont {V.~F.}\ \bibnamefont {Hagh}}, \bibinfo {author} {\bibfnamefont {C.~I.}\ \bibnamefont {Ip}},\ and\ \bibinfo {author} {\bibfnamefont {S.~R.}\ \bibnamefont {Nagel}},\ }\bibfield  {title} {\bibinfo {title} {Competition between energy and dynamics in memory formation},\ }\href@noop {} {\bibfield  {journal} {\bibinfo  {journal} {Phys. Rev. Lett.}\ }\textbf {\bibinfo {volume} {130}},\ \bibinfo {pages} {197201} (\bibinfo {year} {2023})}\BibitemShut {NoStop}%
\bibitem [{\citenamefont {Keim}\ and\ \citenamefont {Arratia}(2013)}]{keim2013yielding}%
  \BibitemOpen
  \bibfield  {author} {\bibinfo {author} {\bibfnamefont {N.~C.}\ \bibnamefont {Keim}}\ and\ \bibinfo {author} {\bibfnamefont {P.~E.}\ \bibnamefont {Arratia}},\ }\bibfield  {title} {\bibinfo {title} {Yielding and microstructure in a 2d jammed material under shear deformation},\ }\href@noop {} {\bibfield  {journal} {\bibinfo  {journal} {Soft Matter}\ }\textbf {\bibinfo {volume} {9}},\ \bibinfo {pages} {6222} (\bibinfo {year} {2013})}\BibitemShut {NoStop}%
\bibitem [{\citenamefont {Regev}\ \emph {et~al.}(2013)\citenamefont {Regev}, \citenamefont {Lookman},\ and\ \citenamefont {Reichhardt}}]{regev2013onset}%
  \BibitemOpen
  \bibfield  {author} {\bibinfo {author} {\bibfnamefont {I.}~\bibnamefont {Regev}}, \bibinfo {author} {\bibfnamefont {T.}~\bibnamefont {Lookman}},\ and\ \bibinfo {author} {\bibfnamefont {C.}~\bibnamefont {Reichhardt}},\ }\bibfield  {title} {\bibinfo {title} {Onset of irreversibility and chaos in amorphous solids under periodic shear},\ }\href@noop {} {\bibfield  {journal} {\bibinfo  {journal} {Phys. Rev. E}\ }\textbf {\bibinfo {volume} {88}},\ \bibinfo {pages} {062401} (\bibinfo {year} {2013})}\BibitemShut {NoStop}%
\bibitem [{\citenamefont {Fiocco}\ \emph {et~al.}(2013)\citenamefont {Fiocco}, \citenamefont {Foffi},\ and\ \citenamefont {Sastry}}]{fiocco2013oscillatory}%
  \BibitemOpen
  \bibfield  {author} {\bibinfo {author} {\bibfnamefont {D.}~\bibnamefont {Fiocco}}, \bibinfo {author} {\bibfnamefont {G.}~\bibnamefont {Foffi}},\ and\ \bibinfo {author} {\bibfnamefont {S.}~\bibnamefont {Sastry}},\ }\bibfield  {title} {\bibinfo {title} {Oscillatory athermal quasistatic deformation of a model glass},\ }\href@noop {} {\bibfield  {journal} {\bibinfo  {journal} {Phys. Rev. E}\ }\textbf {\bibinfo {volume} {88}},\ \bibinfo {pages} {020301} (\bibinfo {year} {2013})}\BibitemShut {NoStop}%
\bibitem [{\citenamefont {Perchikov}\ and\ \citenamefont {Bouchbinder}(2014)}]{perchikov2014variable}%
  \BibitemOpen
  \bibfield  {author} {\bibinfo {author} {\bibfnamefont {N.}~\bibnamefont {Perchikov}}\ and\ \bibinfo {author} {\bibfnamefont {E.}~\bibnamefont {Bouchbinder}},\ }\bibfield  {title} {\bibinfo {title} {Variable-amplitude oscillatory shear response of amorphous materials},\ }\href@noop {} {\bibfield  {journal} {\bibinfo  {journal} {Phys. Rev. E}\ }\textbf {\bibinfo {volume} {89}},\ \bibinfo {pages} {062307} (\bibinfo {year} {2014})}\BibitemShut {NoStop}%
\bibitem [{\citenamefont {Fiocco}\ \emph {et~al.}(2014)\citenamefont {Fiocco}, \citenamefont {Foffi},\ and\ \citenamefont {Sastry}}]{fiocco2014encoding}%
  \BibitemOpen
  \bibfield  {author} {\bibinfo {author} {\bibfnamefont {D.}~\bibnamefont {Fiocco}}, \bibinfo {author} {\bibfnamefont {G.}~\bibnamefont {Foffi}},\ and\ \bibinfo {author} {\bibfnamefont {S.}~\bibnamefont {Sastry}},\ }\bibfield  {title} {\bibinfo {title} {Encoding of memory in sheared amorphous solids},\ }\href@noop {} {\bibfield  {journal} {\bibinfo  {journal} {Phys. Rev. Lett.}\ }\textbf {\bibinfo {volume} {112}},\ \bibinfo {pages} {025702} (\bibinfo {year} {2014})}\BibitemShut {NoStop}%
\bibitem [{\citenamefont {Royer}\ and\ \citenamefont {Chaikin}(2015)}]{royer2015precisely}%
  \BibitemOpen
  \bibfield  {author} {\bibinfo {author} {\bibfnamefont {J.~R.}\ \bibnamefont {Royer}}\ and\ \bibinfo {author} {\bibfnamefont {P.~M.}\ \bibnamefont {Chaikin}},\ }\bibfield  {title} {\bibinfo {title} {{Precisely cyclic sand: Self-organization of periodically sheared frictional grains.}},\ }\href@noop {} {\bibfield  {journal} {\bibinfo  {journal} {Proc. Natl. Acad. Sci.}\ }\textbf {\bibinfo {volume} {112}},\ \bibinfo {pages} {49} (\bibinfo {year} {2015})}\BibitemShut {NoStop}%
\bibitem [{\citenamefont {Lavrentovich}\ \emph {et~al.}(2017)\citenamefont {Lavrentovich}, \citenamefont {Liu},\ and\ \citenamefont {Nagel}}]{lavrentovich2017period}%
  \BibitemOpen
  \bibfield  {author} {\bibinfo {author} {\bibfnamefont {M.~O.}\ \bibnamefont {Lavrentovich}}, \bibinfo {author} {\bibfnamefont {A.~J.}\ \bibnamefont {Liu}},\ and\ \bibinfo {author} {\bibfnamefont {S.~R.}\ \bibnamefont {Nagel}},\ }\bibfield  {title} {\bibinfo {title} {Period proliferation in periodic states in cyclically sheared jammed solids},\ }\href@noop {} {\bibfield  {journal} {\bibinfo  {journal} {Phys. Rev. E}\ }\textbf {\bibinfo {volume} {96}},\ \bibinfo {pages} {020101} (\bibinfo {year} {2017})}\BibitemShut {NoStop}%
\bibitem [{\citenamefont {Lindeman}\ and\ \citenamefont {Nagel}(2021)}]{lindeman2021multiple}%
  \BibitemOpen
  \bibfield  {author} {\bibinfo {author} {\bibfnamefont {C.~W.}\ \bibnamefont {Lindeman}}\ and\ \bibinfo {author} {\bibfnamefont {S.~R.}\ \bibnamefont {Nagel}},\ }\bibfield  {title} {\bibinfo {title} {Multiple memory formation in glassy landscapes},\ }\href@noop {} {\bibfield  {journal} {\bibinfo  {journal} {Science Advances}\ }\textbf {\bibinfo {volume} {7}},\ \bibinfo {pages} {eabg7133} (\bibinfo {year} {2021})}\BibitemShut {NoStop}%
\bibitem [{\citenamefont {Maloney}\ and\ \citenamefont {Lacks}(2006)}]{maloney2006energy}%
  \BibitemOpen
  \bibfield  {author} {\bibinfo {author} {\bibfnamefont {C.~E.}\ \bibnamefont {Maloney}}\ and\ \bibinfo {author} {\bibfnamefont {D.~J.}\ \bibnamefont {Lacks}},\ }\bibfield  {title} {\bibinfo {title} {Energy barrier scalings in driven systems},\ }\href@noop {} {\bibfield  {journal} {\bibinfo  {journal} {Phys. Rev. E}\ }\textbf {\bibinfo {volume} {73}},\ \bibinfo {pages} {061106} (\bibinfo {year} {2006})}\BibitemShut {NoStop}%
\end{thebibliography}
\end{document}